\documentclass[12pt]{iopart}

\usepackage{graphicx}
\usepackage{iopams}
\usepackage[dvips]{hyperref}

\def\ket#1{\vert#1\rangle}
\def\bra#1{\langle #1\vert}

\def\ketbra#1#2{\vert#1\rangle\langle#2\vert}
\def\hc{{\ensuremath{\textrm{\,\it h.\,c.}}}}
\newcommand{\rvert}{\vert}
\newcommand{\lvert}{\vert}

\def\1{\mathchoice{\rm 1\mskip-4.2mu l}{\rm 1\mskip-4.2mu l}{\rm
        1\mskip-4.6mu l}{\rm 1\mskip-5.2mu l}}

\begin{document}
\title[New Aspects of Geometric Phases in Experiments with polarized Neutrons]{New Aspects of Geometric Phases in Experiments with polarized Neutrons}

\author{S. Sponar$^1$, J. Klepp$^{2}$, K. Durstberger-Rennhofer$^1$, R. Loidl$^1$, S. Filipp$^3$,  M. Lettner$^4$, R. A. Bertlmann$^2$, G. Badurek$^1$, H.
Rauch$^{1,5}$ and Y. Hasegawa$^{1}$  }

\address{$^1$ Atominstitut der \"{O}sterreichischen Universit\"{a}ten, TU-Wien, 1020 Vienna, Stadionallee 2, Austria}
\address{$^2$ Faculty of Physics, University of Vienna, Boltzmanngasse 5, 1090 Vienna, Austria}
\address{$^3$ Department of Physics, ETH Z\"{u}rich, Schafmattstr. 16,
8093 Z\"{u}rich, Switzerland}
\address{$^4$ Max-Planck-Institut f\"{u}r Quantenoptik, Hans-Kopfermann-Stra\ss e 1, 85748 Garching, Germany}
\address{$^5$ Institut Laue-Langevin,
B.P. 156, 38042 Grenoble Cedex 9, France}

\ead{sponar@ati.ac.at}

\begin{abstract}
Geometric phase phenomena in single neutrons have been observed in polarimeter and interferometer experiments. Interacting with static and time dependent magnetic fields, the state vectors acquire a geometric phase tied to the evolution within spin subspace. In a polarimeter experiment the non-additivity of quantum phases for mixed spin input states is observed. In a Si perfect-crystal interferometer experiment appearance of geometric phases, induced by interaction with an oscillating magnetic field, is verified. The total system is characterized by an entangled state, consisting of neutron and radiation fields, governed by a Jaynes-Cummings Hamiltonian. In addition, the influence of the geometric phase on a Bell measurement, expressed by the Clauser-Horne-Shimony-Holt (CHSH) inequality, is studied. It is demonstrated that the effect of geometric phase can be balanced by an appropriate change of Bell angles.

\end{abstract}
 \submitto{\JPA} \maketitle

\section{Introduction}

The total phase acquired during an evolution of a quantum system
generally consists of two components: the usual dynamical phase
$\phi_{{d}}$ and the geometric phase $\phi_{{g}}$. The
dynamical phase, which depends on the dynamical properties, like
energy or time is
given by $\phi_{{d}}=-1/\hbar\int H(t) dt $. The  peculiarity of
the geometric phase lies in the fact that it does not depend on the
dynamics of the system, but purely on the evolution path of the
state. Considering a spin $\frac{1}{2}$ system, the geometric phase
is given by minus half the solid angle ($\phi_{{g}}=-\Omega/2$) of the curve
traced out. Since its discovery by M.\,V.\,Berry in
1984 \cite{BerryProcRSocLondA1984} the topological concept has been
widely expanded and has undergone several generalizations.

The first experimental evidence of an adiabatic and cyclic geometric phase, commonly called Berry phase, was achieved with photons in 1986 \cite{TomitaPRL1986}
and later with neutrons \cite{BitterPRL1987}. Non-adiabatic \cite{AharonovPRL1987} and non-cyclic \cite{SamuelPRL1988} geometric phases as well as the
off-diagonal case, where initial and final states are mutually orthogonal \cite{ManiniPRL2000}, have been considered.
In addition to an early approach by Uhlmann
\cite{UhlmannLettMathPhys1991}, an alternative concept of geometric phase for mixed
input states based on interferometry was developed by
Sj\"{o}qvist \emph{et al.} \cite{SjoeqvistPRL2000}.
Here, each eigenvector of the initial
density matrix independently acquires a geometric phase.
The total mixed state phase is a weighted average of the
individual phase factors.
This concept is of great significance for all
experimental situations or technical applications
in which pure state theories oversimplify the description.
Theoretical predictions have
been tested using NMR and single-photon
interferometry \cite{DuPRL2003,EricssonPRL2005}.
The idea has also been extended to the off-diagonal case \cite{FilippPRL2003,FilippPRA2003}.

Neutron interferometry \cite{RauchWerner2000,Rauch-pla74} provides a powerful tool for investigations of quantum phenomena. Particularly in the field of geometric phases, where the spatial as well as the spinor evolution leads to geometric phases: In the spatial case the two-dimensional Hilbert space is spanned by the two possible paths in the interferometer. It has been experimentally verified that a geometric phases for cyclic \cite{hasegawaPRA1996}, as well as non-cyclic evolutions \cite{FilippPRA2005}, can be induced. In the case of spinor evolution, where the geometric phase is generated in spin subspace, the spinor rotations are carried out independently in each sub-beam, due to the macroscopic separation of the partial beams in the interferometer \cite{waghPRL1998}. Geometric phase effects are observed when the two sub-beams are recombined at the third plate of the interferometer followed by a spin analysis. For instance in \cite{Allman-pra97}, spin flippers in both beams clearly demarcate the separate contributions of the dynamical and geometric phase acquired in the spin subspace.

The geometric phase in a single-particle system has been studied widely over the past two and a half decades. Nevertheless its effect on
entangled quantum systems is less investigated. The Berry phase is an excellent candidate to be utilized for logic gate operations in
quantum communication \cite{NielsenCuang2000}, due to its robustness against noise. This has been tested recently using superconducting qubits
\cite{LeekScience2007}, and trapped polarized ultracold neutrons \cite{FilippPRL2009}. Entanglement is the basis for quantum communication and quantum
information processing. Therefore studies on systems combing both quantum phenomena, the geometric phase and quantum entanglement, are of great
importance \cite{BertlmannPRA2004,SjoeqvistPRA2000,tongJPhysA2003}. In the case of neutrons, entanglement is achieved between different degrees of freedom and not between different particles. Using neutron
interferometry, with spin polarized neutrons, single-particle entanglement between the spinor and the spatial part of
the neutron wave function \cite{HasegawaNature2003}, as well as full tomographic state analysis \cite{HasegawaPRA2007}, have already been
accomplished.

In this paper we report on miscellaneous geometric phase phenomena in neutron polarimetry as well as interferometry. In Section \ref{sec:polarimetry}
polarimetric measurements of noncyclic
geometric, dynamical and general phases are presented. In particular, our experiment
demonstrates that the geometric and dynamical mixed state phases
$\Phi_{{{g}}}$ and $\Phi_{{{d}}}$, resulting from separate
measurements, are not additive \cite{SinghPRA2003} in the sense that
the total phase resulting from a single, cumulative, measurement
differs from $\Phi_{{{g}}}+\Phi_{{{d}}}$
\cite{KleppPRL2008,KleppAIPProc2009}. Furthermore, we report on observation and manipulation of the
geometric phase generated in one of the Hilbert spaces in a
spin-path entangled single neutron system, namely the spin subspace.
Section \ref{sec:geophase} focuses on the geometric phase
generation, due to time-dependent interaction with a radio-frequency (rf) field. Here the system is characterized by an
entangled state, consisting of neutron and radiation field, governed
by a Jaynes-Cummings Hamiltonian. In Section
\ref{sec:geophaseinentangledsystems} the influence of the geometric
phase on a spin-path entangled single neutron system is described.
We demonstrate in detail how the geometric phase affects the Bell
angle settings, required for a violation of a Bell-like inequality
in the CHSH formalism.

\section{Experimental observation of non-additivity of mixed-state phases}\label{sec:polarimetry}

\begin{figure}
\begin{center}
\scalebox{0.47}{
\includegraphics{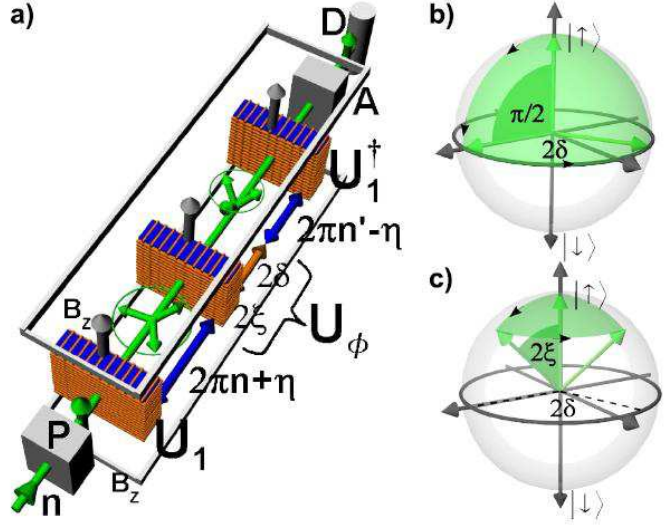}}
\caption{a) Sketch of neutron polarimetry setup for phase measurement with overall guide field $B_z$, polarizer $P$, three DC-coils to implement unitary operations $U_1$, $U^{\dagger}_1$, $U_\phi$, analyzer $A$ and detector $D$. Greek letters denote polarization rotation angles. Shifting the second coil induces an additional dynamical phase $\eta/2$ resulting in spin interference. Evolution of the $|\!\Uparrow\rangle$ state on the Bloch-sphere induced by $U_\phi$, associated to: b) Purely (noncyclic) geometric
phase ($2\xi=\pi/2$). c) Combinations of dynamical and geometric phase on the Bloch sphere ($0<2\xi<\pi/2$).}\label{fig:polsetup}
\end{center}
\end{figure}

\subsection{Neutron polarimeter scheme for phase measurement}

Consider the experimental setup shown in Fig. \ref{fig:polsetup}.
The polarizer $P$ prepares the beam in the $|\!\Uparrow\rangle$
spin state. Subsequently, a coil carries out a $\pi/2$-rotation (U$_1$),
creating a coherent superposition
$1/\sqrt 2\,(|\Uparrow\rangle-i|\Downarrow\rangle)$ of spin eigenstates
that acquire opposite dynamical phase due to Zeeman splitting within
the field B$_z$. Alternatively, one could say that the polarization
vector $\vec r\,'$ rotates in the $x,y$ plane after U$_1$. The
second coil and some arbitrarily chosen propagation distance within
B$_z$ implement a spin evolution U$_\phi$ for both eigenstates and
thereby induces a pure state Pancharatnam (total) phase $\phi$
\cite{Pancharatnam1956}. The third coil (U$_1^{\dagger}$) carries
out a $-\pi/2$-rotation in order to observe spin interference in the detector $D$ after the analyzer $A$ (both, $P$
and $A$ project the spin towards
the $+z$ direction). To obtain these
interferences a phase shift $\eta$ is implemented by linear translation of the second
coil. It was first stated in \cite{WaghPLA1995} that with such an
apparatus one can obtain phases $\phi$ between spin eigenstates of
neutrons, induced by a SU(2) transformation
\begin{eqnarray}\label{transformation}
\!\!\!\!\!\!\!\!\!\!\!\!\!\!\!\!\!\!\!\!\!\!\!\!\!\!\!\!\!\!\!{U}_{\phi}(\xi,\delta,\zeta)=e^{i\delta}\cos\xi|
\!\Uparrow\rangle\langle\Uparrow\!|-e^{-i\zeta}\sin\xi|\!\Uparrow\rangle\langle\Downarrow\!|
+e^{i\zeta}\sin\xi|\!\Downarrow\rangle\langle
\Uparrow\!|+e^{-i\delta}\cos\xi|\!\Downarrow\rangle\langle\Downarrow\!|.
\end{eqnarray}
Equation (\ref{transformation}) describes a general evolution of the
system within static magnetic fields. The resulting total phase
$\phi=\arg\langle\Uparrow\!|{U}_{\phi}|\!\Uparrow\rangle=\delta$ can
be written as a function of the maximum I$_{{max}}$ and minimum
I$_{{min}}$ intensity of the oscillations, exhibited by applying the phase shift $\eta$. The intensity is proportional to
$\cos^2\xi\cos^2\delta+\sin^2\xi \cos^2(\zeta-\eta)$. This only
depends on the SU(2) parameters $\xi$, $\delta$ and $\zeta$ - set by
choosing the spin rotation angles in the second coil and the
additional propagation distance within the guide field B$_z$,
respectively.

A neutron beam with incident purity $r\,'\!=\!|\vec r\,'|$ along the
$+z$-axis ($\vec r\,'\!=\!(0,0,r')$) is described by the density
operator $\rho_{{{in}}}(r)=1/2(\1+r'\sigma_z)$. For mixed input
states, $0\leq r'<1$. In this case \cite{LarssonPLA2003} we find the
intensity oscillations to be proportional to
\begin{eqnarray}\label{mixedstateintensity}
{I}^{\rho}=&& \frac{1-r'}{2}+r'\left(\cos^2\xi\cos^2\delta+\sin^2\xi
\cos^2(\zeta-\eta)\right).\quad
\end{eqnarray}
Considering again the maxima and minima of the intensity oscillations, one
obtains the mixed state phase
\begin{eqnarray}
\!\!\!\!\!\!\!\!\!\!\!\!\!\!\!\Phi(r')&=&
\arccos\sqrt{\frac{[{I}^{\rho}_{{ min}}/{I}^{\rho}_{{
n}}-1/2(1-r')]/r'}
    {r'[1/2(1+r')-{I}^{\rho}_{{ max}}/{I}^{\rho}_{{n}}]+
    [{I}^{\rho}_{{ min}}/{I}^{\rho}_{{ n}}-1/2(1-r')]/r'}}
\label{eq:MixedStatePhInTermsOfImaxImin}
\end{eqnarray}
with a normalization factor ${I}^{\rho}_{{
n}}=2{I}^{\rho}_0/(1+r')$. I$^{\rho}_0$ is the intensity measured with
U$_\phi=\1$.

Generally, the noncyclic geometric phase is given by
$\phi_g=-\Omega/2$, where $\Omega$ is the solid angle enclosed by an
evolution path and its shortest geodesic closure on the Bloch sphere
\cite{SamuelPRL1988}: $\phi_g$ and the total phase $\phi$ are
related to the path by the polar and azimuthal angles $2\xi$ and
$2\delta$ respectively, so that the pure state geometric phase can be written as
\begin{eqnarray}\label{eq:gamma(delta,xi)}
    \phi_{{{g}}}
=\phi- \phi_{{{d}}}=\delta[1-\cos{(2\xi)}].
\end{eqnarray}
$\phi_{{{d}}}$ is the dynamical phase. By proper choice of $2\xi$
and $2\delta$, U$_\phi$ can be set to generate purely geometric,
purely dynamical, or arbitrary combinations of both phases, e.g. in Figs.\,\ref{fig:polsetup}b and c: For instance, we can
choose to induce a purely geometric phase by selecting $2\xi$ to be
equal to $\pi/2$.

The theoretical prediction for the mixed state phase is
\cite{SjoeqvistPRL2000,LarssonPLA2003}
\begin{eqnarray}
\Phi&=&\arctan\left(r'\tan\delta\right)\label{eq:MixedStatePhase}
\end{eqnarray}
Note that Eq.\,(\ref{eq:MixedStatePhase}) only depends on the
parameter $\delta$ and the purity $r'$. Again, as can be seen also
from Eq. (\ref{eq:gamma(delta,xi)}), the parameter $\xi$ only
determines the portion of dynamical phase $\phi_{{{d}}}$ contained
in the total phase $\phi$.

\subsection{Experiment}

To access Eq. (\ref{eq:MixedStatePhase}) experimentally $r'$ has to
be varied. In addition to the DC current, which effects the
transformation U$_1$, random noise is applied to the first coil,
thereby changing B$_x$ in time. Neutrons, which are part of the
ensemble $\rho_{{{in}}}(r')$, arrive at different times at the coil
and experience different magnetic field strengths. We are left with
the system in a mixed state $\vec r\!=\!\left(0,-r,0\right)$ where
$r\!<\!1$ \cite{BertlmannPRA2006}.
\begin{figure}
\begin{center}
\scalebox{0.4}{\includegraphics{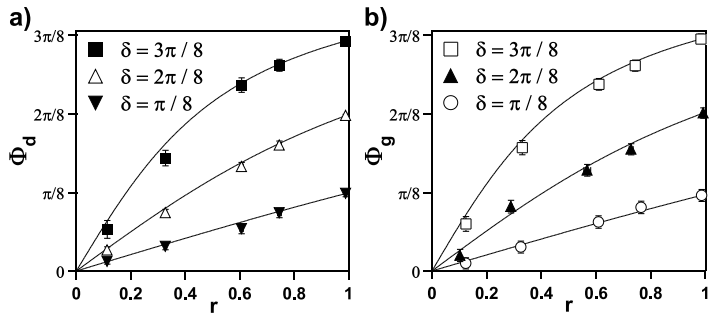}} \caption{Mixed state
phases determined from measured intensity oscillations using
Eq.\,(\ref{eq:MixedStatePhase}). Dynamical phase (a), geometric
phase (b) versus input purity $r$ for three evolution paths, i. e.
three settings of the second coil (angle $2\xi$) and flight distance
after it (angle $2\delta$). The legends indicate evolutions. Solid
lines are theory curves using the rightmost data points as
reference.}\label{fig:phasevisib}
\end{center}
\end{figure}

A neutron beam -- incident from a pyrolytic graphite crystal -- with
a mean wavelength $\lambda\!\approx\!1.98$\,\AA ~and spectral width
$\Delta\lambda/\lambda\!\approx\!0.015$, was polarized up to 99\% by
reflection from a bent Co-Ti supermirror array. The final maximum
intensity was about 150\,counts/s at a beam cross-section of roughly
1\,cm$^2$. A $^3$He gas detector was used. Spin rotations around the
$+x$-axis were implemented by magnetic fields B$_x$ from DC coils on
frames with rectangular profile ($7\!\times\!7\!\times\!2$ cm$^3$).
B$_z$ was realized by two rectangular coils of 150\,cm length in
Helmholtz geometry. Low coil currents of about 2\,A corresponding to
field strengths of up to 1\,mT were required for the spin rotations
and to prevent unwanted depolarization. The noise from a standard
signal generator consisted of random DC offsets varying at a rate of
20\,kHz. The experimental data shown in Fig. \ref{fig:phasevisib}
reproduce well the $r'$-dependence predicted by
Eq.\,(\ref{eq:MixedStatePhase}).

\subsection{Non-additivity}

Our experiment focuses on a special property of the mixed state
phase: its non-additivity. The Sj\"{o}qvist mixed state phase \cite{SjoeqvistPRL2000} is
defined as a weighted average of phase factors rather than one of
phases. So it is true only for pure states that phases accumulated
in separate experiments can be added up to the usual total phase in
the following sense: Suppose a geometric pure state phase  $\phi_{{{g}}}$ is induced in a first, and a dynamic pure state phase $\phi_{{{d}}}$ in a second measurement.
\begin{figure}
\begin{center}
\scalebox{0.6}{\includegraphics{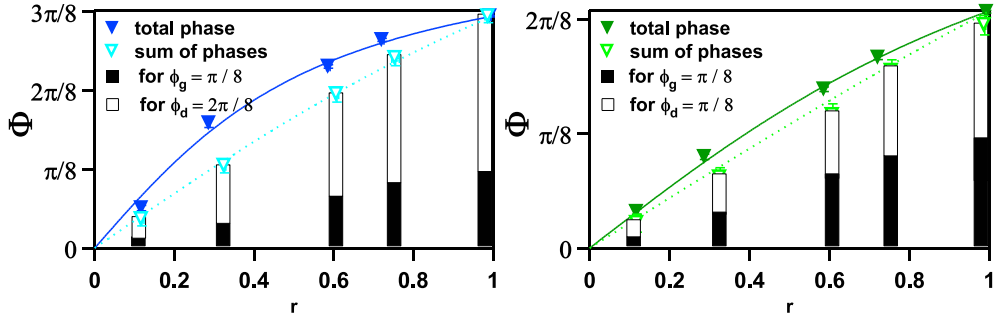}} \caption{Filled
markers: Measured total mixed state phase $\Phi_{{{tot}}}$ versus
purity $r$ for two examples of U$_{{{tot}}}$ associated to the total
pure state phases $\phi_{{{g}}} +\phi_{{{d}}}$ (see text). Open
markers: $\Phi_{{{g}}}+\Phi_{{{d}}}$ as calculated from measured
data. Filled (empty) bars show measured mixed-state geometric
(dynamical) phases. The solid and dotted theory curves assume either
non-additivity or additivity, respectively.}\label{fig:results}
\end{center}
\end{figure}
Applying (\ref{eq:gamma(delta,xi)}) we can also choose a combination
of angles $2\xi$ and $2\delta$ leading to a transformation
U$_{{{tot}}}$, so that we measure the total pure state phase
$\phi_{{{g}}}+\phi_{{{d}}}$ in a third experiment. However, the
result of this latter experiment for the system initially in a mixed
state is $\Phi_{{{tot}}}(r) =\arctan{\left[r\tan(\phi_{{{g}}}
+\phi_{{{d}}})\right]}$. The total phase is \emph{not} equal to
$\Phi_{{{g}}}(r) +\Phi_{{{d}}}(r)$, with $\Phi_{{{g}}}(r)
=\arctan{(r\tan\phi_{{{g}}})}$ and $\Phi_{{{d}}}(r)
=\arctan{(r\tan\phi_{{{d}}})}$. Two examples of data confirming this
prediction are shown in Fig. \ref{fig:results}.

\section{Geometric phase generation in an oscillating magnetic field}\label{sec:geophase}

The evolution of a system consisting of neutron, static magnetic field and quantized rf-field can be described by a photon-neutron state vector,
which is an eigenvector of a Jaynes-Cummings (J-C) Hamiltonian \cite{JaynesProcIEEE1963,ShoreJModOpt1993}, adopted for this particular physical configuration
\cite{MuskatPRL1987}. Since two rf-fields (the reason for the second rf-field is explained in Section \ref{sec:experiment}), operating at frequencies $\omega$ and $\omega/2$, are involved in the actual experiment,
the modified corresponding J-C Hamiltonian is denoted as
\begin{equation}
\mathcal H_{\rm{J-C}}=-\frac{\hbar^2}{2m}\nabla^2-\mu B_0(\textbf{r}) \sigma_z+\hbar(\omega a_{\omega}^\dagger a_{\omega}+\frac{\omega}{2}
a_{\omega/2}^\dagger a_{\omega/2})
\end{equation}
\begin{equation}\label{eq:J-CHamiltonian}
\!\!\!\!\!+\mu\Bigg(\frac{B^{(\omega)}_1(\textbf{r})}{\sqrt{N_{\omega}}}(
a_{\omega}^\dagger\widetilde\sigma+\hc)+\frac{B^{(\omega/2)}_1(\textbf{r})}{\sqrt{N_{\omega/2}}}(
a_{\omega/2}^\dagger\widetilde\sigma+\hc)\Bigg).
\end{equation}
with $\widetilde\sigma=\frac{1}{2}(\sigma_x+ i\sigma_y)$. The first term accounts for the kinetic energy of the neutron. The second term leads
to the usual Zeeman splitting of $2\vert\mu\vert B_0$. The third term adds the photon energy of the oscillating fields of frequencies $\omega$
and $\omega/2$, by use of the creation and annihilation operators $a^\dagger$ and $a$. Finally, the last term represents the coupling between
photons and the neutron, where $N_{\omega_j}=\langle a_{\omega_j} ^\dagger a_{\omega_j}\rangle$ represents the mean number of photons with
frequencies $\omega_j$ in the rf-field.

The state vectors of the oscillating fields are represented by coherent states $\ket{\alpha}$, which are eigenstates of $a^\dagger$ and $a$. The
eigenvalues of coherent states are complex numbers, so one can write $a\ket{\alpha}=\alpha\ket{\alpha}=\vert\alpha\vert
e^{i\phi}\ket{\alpha}\rm{ with } \vert\alpha\vert=\sqrt{N}$. Neutrons interacting with electromagnetic quanta are usually described by the
'dressed-particle' formalism \cite{MuskatPRL1987}, in analogy to the dressed-atom concept \cite{CohenTannJPhys1969}
developed nearly two decades before. Using Eq.\,(\ref{eq:J-CHamiltonian}) one can define a total state vector including not only the neutron
system $\ket{\Psi_{\rm{N}}}$, but also the two quantized oscillating magnetic fields:
\begin{equation}
\ket{\Psi_{\rm{tot}}}=\ket{\alpha_\omega}\otimes\ket{\alpha_{\omega/2}}\otimes\ket{\Psi_{\rm{N}}}.
\end{equation}
In a  perfect Si-crystal neutron interferometer the wavefunction behind the first plate, acting as a beam splitter, is a linear superposition of
the sub-beams belonging to the right ($\ket{\textrm{I}}$) and the left path ($\ket{\textrm{II}}$), which are laterally separated by several
centimeters. The sub-beams are recombined at the third crystal plate and the wave function in forward direction then reads as
$\ket{\Psi_{\rm{N}}}\propto\ket{\textrm{I}}+\ket{\textrm{II}}$, where $\ket{\textrm{I}}$ and $\ket{\textrm{II}}$ only differ by an
adjustable phase factor $e^{i\chi}$ ($\chi=-N_{\rm{ps}}b_c\lambda D$, with the atom number density $N_{\rm{ps}}$ in the phase shifter plate, the coherent scattering length $b_c$, the neutron wavelength $\lambda$ and the thickness of the phase shifter plate $D$). By rotating
the plate, $\chi$ can be varied. This yields the well known sinusoidal intensity oscillations of the two beams emerging behind the
interferometer, usually denoted as O- and H-beam \cite{RauchWerner2000}.

\begin{figure}
\begin{center}
{\includegraphics [width=155mm] {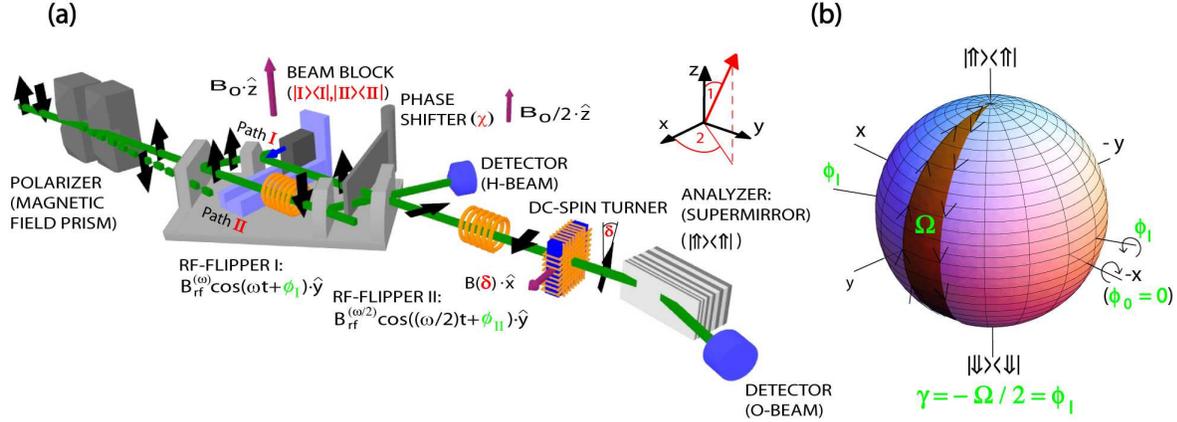}} \caption {(a) The experimental apparatus for observation of geometric phase. The spin state acquires
a geometric phase $\gamma$ during the interaction with the two rf-fields and is flipped twice. Finally, the spin is rotated by an angle
$\delta=\pi/2$ (in the $\hat{\mathbf x}, \hat{\mathbf z}$ plane), by a dc-spin turner, for a polarization analysis and count rate detection. (b)
The Bloch-sphere description depicts the acquired geometric phase given by minus half the solid angle depending on the phase $\phi_{{\rm
I}}$ of the rf-field. The effect of the beam block is explained in Section
\ref{sec:geophaseinentangledsystems}.}\label{fig:setup}
\end{center}
\end{figure}

In our experiment, only the beam in path II is exposed to the rf-field of frequency $\omega$, resulting in a spin flip (see
Fig.\,\ref{fig:setup}\,(a)). Interacting with a time-dependent magnetic field, the total energy of the neutron is no longer conserved after the
spin-flip \cite{alefeldZPhysB1981,GaehlerPLA1987,SummhammerPRA1993,GolubAmJPhys1994,SummhammerPRL1995,GrigorievPRA2004}. Photons of energy $\hbar\omega$ are
exchanged with the rf-field.

The time-dependent entangled state, which emerges from a coherent superposition of $\ket{\rm{I}}$ and $\ket{\rm{II}}$, is expressed as
\begin{eqnarray}\label{eq:totstate}
\ket{\Psi_{\rm{tot}}}=\ket{\alpha_\omega}\otimes\ket{\alpha_{\omega/2}}\otimes\frac{1}{\sqrt{2}}\Big(\ket{\textrm{I}}\otimes\ket{\Uparrow}+e^{i
\omega t}e^{i\chi}\ket{\textrm{II}}\otimes e^{i\phi_{\rm{I}}}\ket{\Downarrow}\Big),
\end{eqnarray}
for a more detailed description of the generation of $\ket{\Psi_{\rm{tot}}}$ see \cite{SponarPRA2008}.

The effect of the first rf-flipper, placed inside the interferometer (path II), is described by the unitary operator $\hat U(\phi_{\rm I})$,
which induces a spinor rotation from $\ket{\Uparrow}$ to $\ket{\Downarrow}$, we denoted $\hat U(\phi_{\rm
I})\ket{\Uparrow}=e^{i\phi_{\rm I}}\ket{\Downarrow}$. The rotation axis encloses an angle $\phi_{\rm I}$ with the $\hat {\mathbf
x}$-direction, in the rotating frame, and is determined by the oscillating magnetic field $B^{(1)}=B_{\rm{rf}}^{(\omega)} \cos(\omega t
+\phi_{\rm{I}})\cdot\hat{\mathbf y}$. Formally one can insert a unity operator, given by $\1= \hat U^\dagger(\phi_0)\hat
U(\phi_0)$, yielding
\begin{eqnarray}
\hat U(\phi_{\rm I})\ket{\Uparrow}&=&\overbrace {\hat U(\phi_{\rm I})\hat U^\dagger(\phi_0)}^{ e^{i\gamma}}\hat
U(\phi_0)\ket{\Uparrow}\hspace{-3.05cm}\underbrace{\hspace{2.3cm}}_{\1}\hspace{0.5cm}\quad =e^{i\gamma}\ket{\Downarrow},
\end{eqnarray}

where $\hat U(\phi_0)$ can be interpreted as a rotation from $\ket{\Uparrow}$ to $\ket{\Downarrow}$, with the $\hat {\mathbf x}$-direction being
the rotation axis ($\phi_0=0$), and $\hat U^\dagger(\phi_0)$ describes a rotation about the same axis back to the initial state
$\ket{\Uparrow}$. Consequently, $\hat U(\phi_{\rm I})\hat U^\dagger(\phi_0)$ can be identified to induce the geometric phase $\gamma$, along
the reversed evolution path characterized by $\phi_0$ ($\ket{\Downarrow}$ to $\ket{\Uparrow}$), followed by another path determined by
$\phi_{\rm I}$ ($\ket{\Uparrow}$ to $\ket{\Downarrow}$), see Fig.\,\ref{fig:setup}\,(b). In the rotating frame of reference \cite{SuterPRL1988} the two semi-great circles
enclose an angle $\phi_{\rm{I}}$ and the solid angle $\Omega=-2\phi_{\rm{I}}$, yielding a pure geometric phase
\begin{equation}\label{eq:GeoPhase}
\gamma=-\Omega/2=\phi_{\rm{I}},
\end{equation}
which is depicted in Fig.\,\ref{fig:setup}\,(b). The entangled state, as described in \cite{BertlmannPRA2004}, is represented by
\begin{equation}\label{eq:stateExp}
\ket{\Psi_{\rm{Exp}}(\gamma)}=\frac{1}{\sqrt{2}}\Big(\ket{\textrm{I}}\otimes\ket{\Uparrow}+\ket{\textrm{II}}\otimes
e^{i\gamma}\ket{\Downarrow}\Big),
\end{equation}
including the geometric phase $\gamma=\phi_{\rm I}$.

\begin{figure}
\begin{center}
{\includegraphics [width=125mm] {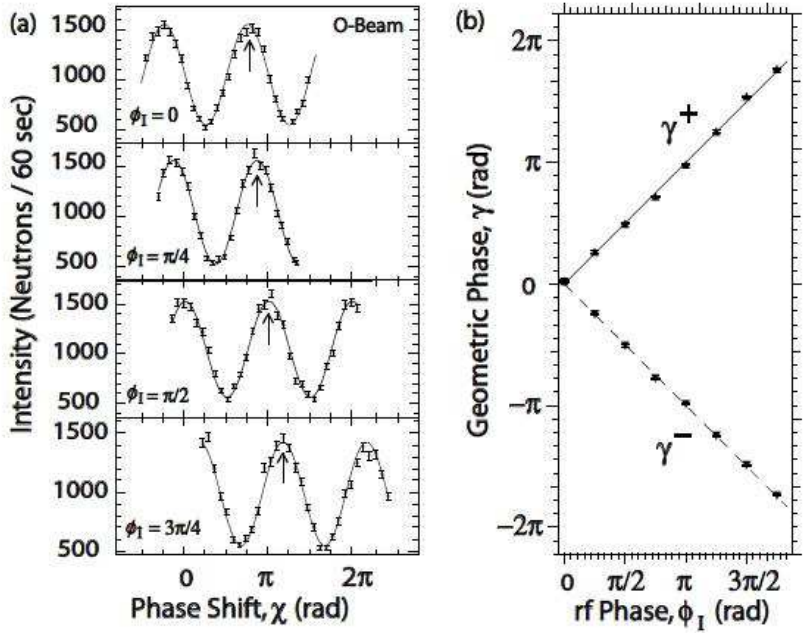} } \caption {(a) Typical interference patterns of the O-beam, when rotating the phase shifter
plate ($\chi$).(b) A phase shift occurs by varying $\phi_{\rm I}$ determining the geometric phase $\gamma$ . The sign of the geometric phase $
\gamma^\pm$ depends on the chosen initial polarization.}\label{fig:oscillations}
\end{center}
\end{figure}

\subsection{Experimental Setup}\label{sec:experiment}

As in a previous experiment \cite{SponarPRA2008}, the spin in path $\ket{\textrm{II}}$ is flipped by a rf-flipper, which requires two magnetic
fields: A static field $B_0\cdot\hat{\mathbf z}$ and a perpendicular oscillating field $B^{(1)}=B_{\rm{rf}}^{(\omega)} \cos(\omega t
+\phi_{\rm{I}})\cdot\hat{\mathbf y}$ satisfying the amplitude and frequency resonance condition
\begin{equation}
B^{(\omega)}_{\rm{rf}}=\frac{\pi\hbar}{\tau\vert\mu \vert}\textrm{ and }  \omega=\frac{2\vert\mu \vert B_0}{\hbar}(1+\frac{B_1^2}{16B_0^2}),
\end{equation}
where $\mu $ is the magnetic moment of the neutron and $\tau$ denotes the time the neutron is exposed to the rf-field.  The second term in
$\omega$ is due to the Bloch-Siegert shift \cite{BlochPR1940}. The oscillating field is produced by a water-cooled rf-coil with a length
of 2\,cm, operating at a frequency of $\omega/2\pi=58$\,kHz. The static field is provided by a uniform magnetic guide field $B_0^{(\omega)}\sim
2$\,mT, produced by a pair of water-cooled Helmholtz coils.

The O-beam passes the second rf-flipper, operating at
$\omega/2\pi=29$\,kHz, which is half the frequency of the first
rf-flipper. The oscillating field is denoted as
$B^{(\omega/2)}_{\rm{rf}} \cos\big((\omega/2) t
+\phi_{\rm{II}}\big)\cdot\hat{\mathbf y}$, and the strength of the
guide field was tuned to $B_0^{(\omega/2)}\sim1\,$mT in order to
satisfy the frequency resonance condition. This flipper compensates
the energy difference between the components from the two interfering paths, by absorption
and emission of photons of energy $E=\hbar\omega/2$. By choosing a
frequency of $\omega/2$ for the second rf-flipper, the
time-dependence of the state vector is eliminated since both components acquire a phase $e^{\pm i\omega/2 (t+T)}$, depending on
the spin orientation. Only a constant phase offset of $e^{\omega
T}$, where $T$ is the  propagation time between the centre of the
first and second flipper coil, remains in the stationary state
vector. This phase contribution,
together with a dynamical phase contribution, resulting from Larmor precession
within the guide field regions $B_0^{(\omega)}$ and
$B_0^{(\omega/2)}$ (pointing in $+\,\hat{\mathbf z}$-direction), are
omitted here because they remain constant during the entire experiment.
Finally, the spin is rotated by an angle $\delta=\pi/2$ (in the
$\hat{\mathbf x}, \hat{\mathbf z}$ plane) with a static field
spin-turner, and analyzed due to the spin dependent reflection
within a Co-Ti multi-layer supermirror along the $\hat{\mathbf
z}$-direction. Intensity oscillations in forward direction (O-beam) are plotted in Fig.\,\ref{fig:oscillations}\,(a).

In a non-dispersive arrangement of the monochromator and the interferometer crystal the angular separation can be used such that only the spin-up (or
spin-down) component fulfils the Bragg-condition at the first interferometer plate (beam splitter). Therefore it is possible to invert the
initial polarization simply by rotating the interferometer by a few seconds of arc, which is expected to lead to an inversion of the geometric
phase. Figure \,\ref{fig:oscillations}(b) shows a plot of the geometric phase $\Delta\gamma^\pm$ versus $\phi_{\rm I}$, with $\phi_{\rm II}=0$.
As expected, the slope $s$ is positive for initial spin up orientation ($s=1.007(8)$), and negative for the spin down case ($s=-0.997(5)$), as predicted in Eq.(\ref{eq:GeoPhase})

\section{Geometric phase effects on a spin-path entangled System} \label{sec:geophaseinentangledsystems}

In this Section the influence of the geometric phase on a Bell measurement \cite{BellPhys1964}, expressed by the Clauser-Horne-Shimony-Holt (CHSH)
\cite{ClauserPRL1969} inequality, as proposed in \cite{BertlmannPRA2004} is discussed . Following the notation given in
\cite{BertlmannPRA2004}, the neutron\char39{}s wavefunction is defined via tensor product of two Hilbert spaces: One Hilbert
space is spanned by two possible paths in the interferometer given by $\ket{\textrm{I}}$ and $\ket{\textrm{II}}$; the other one by spin-up and
spin-down eigenstates, denoted as $\ket{\Uparrow}$ and $\ket{\Downarrow}$, with respect to a quantization axis along a static magnetic field. For
this experiment we focus on the neutron part of Eq.(\ref{eq:totstate}) and omit all phases but the
geometric phase $\gamma$:
\begin{equation}
\ket{\Psi_{\rm{N}}(\gamma)}=\frac{1}{\sqrt{2}}\Big(\ket{\textrm{I}}\otimes\ket{\Uparrow}+ \ket{\textrm{II}}\otimes
e^{i\gamma}\ket{\Downarrow}\Big).
\end{equation}
As in common Bell experiments a joint measurement for spin and path is performed, thereby applying the projection operators for the path
\begin{equation}
\hat P^{\rm{p}}_\pm(\boldsymbol{\alpha})=\ketbra{\pm\boldsymbol{\alpha}}{\pm\boldsymbol{\alpha}},
\end{equation}
with
\begin{equation}
\fl\ket{+\boldsymbol\alpha}=\cos\frac{\alpha_1}{2}\ket{\textrm{I}}+e^{i\alpha_2}\sin\frac{\alpha_1}{2}\ket{\textrm{II}} \quad\rm{and}\quad
\ket{-\boldsymbol\alpha}=-\sin\frac{\alpha_1}{2}\ket{\textrm{I}}+e^{i\alpha_2}\cos\frac{\alpha_1}{2}\ket{\textrm{II}},
\end{equation}
where $\alpha_1$ denotes the polar angle and $\alpha_2$ the azimuthal angle for the path. This is done in analogous manner for the spin subspace with
${\beta_1}$ as the polar angle and ${\beta_2}$ as the azimuthal angle for the spin. Introducing the observables
\begin{equation}
    \hat A^{\rm{p}}(\boldsymbol{\alpha})=\hat P_{+}^{\rm{p}}(\boldsymbol{\alpha})-\hat
    P_{-}^{\rm{p}}(\boldsymbol{\alpha})\quad\rm{and}\quad
    \hat B^{\rm{s}}(\boldsymbol{\beta})=\hat P_{+}^{\rm{s}}(\boldsymbol{\beta})-\hat P_{-}^{\rm{s}}(\boldsymbol{\beta})\
\end{equation}
one can define an expectation value for a joint measurement of spin and path along the directions $\boldsymbol\alpha$ and $\boldsymbol\beta$
\begin{eqnarray}\label{eq:jointMeasurement}
\fl E(\boldsymbol\alpha,\boldsymbol\beta)=\bra{\Psi} \hat A^{\rm{p}}(\boldsymbol\alpha)\otimes \hat
B^{\rm{s}}(\boldsymbol\beta)\lvert\Psi\rangle =-\cos\alpha_1\cos\beta_1-\cos(\alpha_2-\beta_2+\gamma)\sin\alpha_1\sin\beta_1 \\\nonumber
  =-\cos(\alpha_1-\beta_1)\textrm{ for } (\alpha_2-\beta_2)= -\gamma.
\end{eqnarray}

Next, a Bell-like inequality in CHSH-formalism \cite{ClauserPRL1969} is introduced, consisting of four expectation values with the associated
directions $\boldsymbol\alpha$, $\boldsymbol\alpha'$ and $\boldsymbol\beta$, $\boldsymbol\beta'$ for joint measurements of spin and path,
respectively
\begin{equation}\label{eq:S-functionOrg}
S(\boldsymbol\alpha,\boldsymbol\alpha',\boldsymbol\beta,\boldsymbol\beta',\gamma)=\big\vert
E(\boldsymbol\alpha,\boldsymbol\beta)-E(\boldsymbol\alpha,\boldsymbol\beta')
 + E(\boldsymbol\alpha',\boldsymbol\beta)+E(\boldsymbol\alpha',\boldsymbol\beta')\big\vert
\end{equation}
Without loss of generality one angle can be eliminated by setting, e.g., $\boldsymbol\alpha=0$ ($\alpha_1=\alpha_2=0$), which gives
\begin{eqnarray}\label{S-function1}
    \fl S(\boldsymbol\alpha',\boldsymbol\beta,\boldsymbol\beta',\gamma)=\Bigl\lvert -\sin\alpha'_1\Bigl(\cos(\alpha'_2-\beta_2-\gamma)\sin\beta_1
    +\cos(\alpha'_2-\beta'_2-\gamma)\sin\beta'_1\Bigr)\\\nonumber
    -\cos\alpha'_1(\cos\beta_1+\cos\beta'_1)  -\cos\beta_1+\cos\beta'_1\Bigr\rvert.
\end{eqnarray}
The boundary of Eq.(\ref{eq:S-functionOrg}) is given by the value 2 for any noncontextual hidden-variable theories \cite{BasuPLA2001}. Keeping the polar angles $\alpha'_1$, $\beta_1$
and $\beta'_1$ constant at the usual Bell angles $\alpha'_1=\frac{\pi}{2}$, $\beta_1=\frac{\pi}{4}$, $\beta'_1=\frac{3\pi}{4}$ (and azimuthal
parts fixed at $\alpha'_2=\beta_2=\beta'_2=0$) reduces $S$ to
\begin{eqnarray}\label{eq:Sunchganed}
    S(\gamma)= \big\lvert-\sqrt{2}-\sqrt{2}\cos\gamma\big\rvert,
\end{eqnarray}
where the familiar maximum value of $2\sqrt{2}$ is reached for $\gamma=0$. For $\gamma=\pi$ the value of $S$ approaches zero.

\begin{figure}[b]
\begin{center}
\scalebox{0.65}{\includegraphics{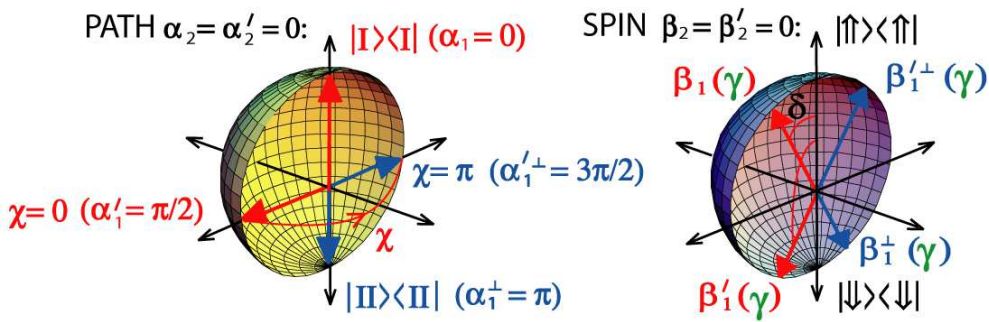}}
\caption{\label{fig:polar}Bloch-sphere description
includes the measurement settings of $\boldsymbol\alpha$ and
$\boldsymbol\beta(\delta)$, determining the projection operators,
used for joint measurement of spin and path. $\boldsymbol\alpha$ is
tuned by a combination of the phase shifter ($\chi$) and the beam
block, and $\boldsymbol\beta$ is adjusted by the angle $\delta$.}
\end{center}
\end{figure}

\subsection{Polar Angle Adjustment}

Here we consider the case when the azimuthal angles are kept
constant, e.g., $\alpha'_2=\beta_2=\beta'_2=0$ ($\alpha_2=0)$, which is depicted in Fig.\,\ref{fig:polar}
The $S$-function reads as
\begin{eqnarray}\label{eq:S-functionAzFixed}
\fl S(\alpha'_1,\beta_1,\beta'_1,\gamma)=\Bigl\lvert -\sin\alpha'_1\Bigl(\cos\gamma\sin\beta_1+\cos\gamma\sin\beta'_1\Bigr)
-\cos\alpha'_1(\cos\beta_1+\cos\beta'_1) \\\nonumber
  -\cos\beta_1+\cos\beta'_1\Bigr\rvert.
\end{eqnarray}

\begin{figure}[t]
\begin{center}
\scalebox{0.29}{\includegraphics{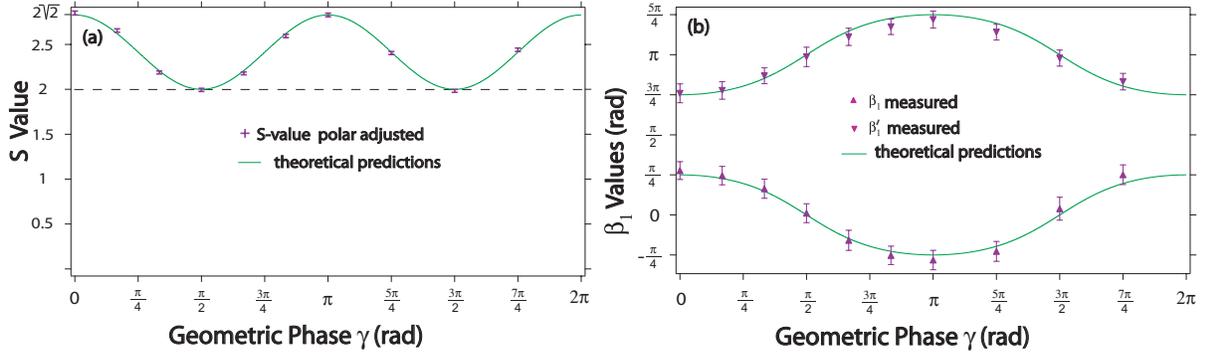}}
\caption{\label{fig:ResultsPolAdjust}  (a) \emph{Polar adjusted}
$S$-values versus geometric phase $\gamma$ with adapted Bell angles
($\beta_1$ and $\beta'_1$) according to the geometric phase
$\gamma$. (b) the corresponding modified Bell angles are plotted
versus the geometric phase $\gamma$.}
\end{center}
\end{figure}

The polar Bell angles $\beta_1$, $\beta'_1$ and $\alpha'_1$ ($\alpha_1=0$), yielding a maximum $S$-value, can be determined, with respect to the
geometric phase $\gamma$, by calculating the partial derivatives (the extremum condition) of $S$ in Eq.(\ref{eq:S-functionAzFixed})(see
\cite{BertlmannPRA2004} for more elaborated deduction): The solutions are given by
\begin{eqnarray}\label{eq:Bell-angles}
\beta_1 = \arctan\big(\cos\gamma\big)\label{eq:Cond1},\quad \beta'_1
= \pi - \beta_1\label{eq:Cond2}\quad\rm{and}\quad \alpha'_1 =
\frac{\pi}{2},
\end{eqnarray}
which are plotted in Fig.\,\ref{fig:ResultsPolAdjust}\,(b) (denoted as
theoretical predictions). With these angles the maximal $S$ decreases from $S=2\sqrt2$
for $\gamma : 0\rightarrow\frac{\pi}{2}$ and touches at $\gamma =
\frac{\pi}{2}$ even the limit of the CHSH inequality $S=2$. Within the interval $\gamma : \frac{\pi}{2}\rightarrow\pi$ the value of $S$ increases again and returns to the familiar value $S=2\sqrt2$ at $\gamma=\pi$.

Experimentally, the probabilities of joint (projective) measurements
are proportional to the following count rates $N_{ij}$ with $(i,j=+,-)$, detected after path
($\boldsymbol\alpha$) and spin ($\boldsymbol\beta$) manipulation:
\begin{eqnarray}\label{expectation_value_exp}
  E(\boldsymbol\alpha,\boldsymbol\beta)=
\frac{N_{++}(\boldsymbol\alpha,\boldsymbol\beta)-N_{+-}(\boldsymbol\alpha,\boldsymbol\beta)-N_{-+}(\boldsymbol\alpha,\boldsymbol\beta)
+N_{--}(\boldsymbol\alpha,\boldsymbol\beta)}
    {N_{++}(\boldsymbol\alpha,\boldsymbol\beta)+N_{+-}(\boldsymbol\alpha,\boldsymbol\beta)+N_{-+}(\boldsymbol\alpha,\boldsymbol\beta)
    +N_{--}(\boldsymbol\alpha,\boldsymbol\beta)},
\end{eqnarray}
with for example
\begin{eqnarray}
   \fl N_{++}(\boldsymbol\alpha,\boldsymbol\beta)=N_{++}\big(\boldsymbol\alpha,(\beta_1,0)\big)\propto\langle
    \Psi_{\rm{N}}(\gamma)\rvert \hat P_+^{\rm{p}}(\boldsymbol\alpha)\otimes
    \hat
    P_+^{\textrm{s}}(\beta_1,0)\lvert\Psi_{\rm{N}}(\gamma)\rangle
\end{eqnarray}
In the case of $N_{+-}(\boldsymbol\alpha,\boldsymbol\beta)$ the count rate
is given by $N_{++}\big(\boldsymbol\alpha,(\beta_1^\bot,0)\big)$,
where $\beta_1^\bot=\beta_1+\pi$. The procedure is applied to the
count rates  $N_{+-}(\boldsymbol\alpha,\boldsymbol\beta)$ and
$N_{--}(\boldsymbol\alpha,\boldsymbol\beta)$.
With these expectation values S can be calculated as defined in Eq.(\ref{eq:S-functionOrg}).

Projective measurements are performed on parallel planes defined by
$\alpha_2=\alpha'_2=\beta_2=\beta'_2=0$.
For the path measurement the directions are given by
$\boldsymbol\alpha:\alpha_1=0, \alpha_2=0$ , and
$\boldsymbol\alpha': \alpha'_1=\pi/2,\,\alpha'_2=0$.

The angle $\boldsymbol\alpha$, which corresponds to $+\,\hat{\mathbf
z}$ (and $-\,\hat{\mathbf z}$ for
$\alpha_1^\bot=\alpha_1+\pi=\pi,\alpha_2=0$) is achieved by the use
of a beam block which is inserted to stop beam II (I) in order to
measure along $+\,\hat{\mathbf z}$ (and $-\,\hat{\mathbf z}$). The
corresponding operators are given by
\begin{eqnarray}\label{eq:pol1}
\hat P^{\rm{p}}_{+z}({\alpha_1=0, \alpha_2=0})&=&\ketbra{\textrm{I}}{\textrm{I}}\nonumber\\
\hat P^{\rm{p}}_{-z}({\alpha_1^\bot=\pi,
\alpha_2=0})&=&\ketbra{\textrm{II}}{\textrm{II}},
\end{eqnarray}

The angle $\boldsymbol\alpha'$ is set by a superposition of equal
portions of $\ket{\textrm{I}}$ and $\ket{\textrm{II}}$, represented
on the equator of the Bloch sphere. The interferograms are achieved
by a rotation of the phase shifter plate, associated with a
variation of the path phase $\chi$. All path scans are repeated at
different values of the spin analysis direction $\delta$ in order to
determine $\beta_1$ and $\beta_1^\prime$ for a maximal violation of
the Bell-like CHSH inequality. The projective measurement for
$\alpha'_1=\pi/2,\alpha'_2=0$ corresponds to a phase shifter
position of $\chi$=~0 (and
$\alpha'_1$ $^\bot$=~$\alpha'_1+\pi=3\pi/2,\alpha'_2=~0$ to
$\chi=\pi$). Projection operators read as
\begin{eqnarray}\label{eq:pol2}
&\hat P^{\rm{p}}_{+x}({\alpha'_1=\frac{\pi}{2}, \alpha'_2=0})=\frac{1}{2}\Big(\big(\ket{\textrm{I}}+\ket{\textrm{II}}\big)\big(\bra{\textrm{I}}+\bra{\textrm{II}}\big)\Big)\\
&\hat P^{\rm{p}}_{-x}({\alpha'^{\bot}_1=\frac{3\pi}{2},
\alpha'_2=0})=
\frac{1}{2}\Big(\big(\ket{\textrm{I}}-\ket{\textrm{II}}\big)\big(\bra{\textrm{I}}-\bra{\textrm{II}}\big)\Big).
\end{eqnarray}
Using the measurement curves from Eq.(\ref{eq:pol1}) and
Eq.(\ref{eq:pol2}), the $S$-value is calculated according to
Eq.(\ref{eq:S-functionOrg}) as a function of the parameters $\beta_1$ and
$\beta_1'$, which are varied independently. The local maximum of the
$S(\beta_1',\beta_1)$ is determined numerically and plotted in
Fig.\,\ref{fig:ResultsPolAdjust}\,(a), with the corresponding values
for $\beta_1$ and $\beta_1'$ in Fig.\,\ref{fig:ResultsPolAdjust}\,(b).

\subsection{\label{sec:level3}Azimuthal Angle Adjustment}

\begin{figure}[b]
\begin{center}
\scalebox{0.65}{\includegraphics{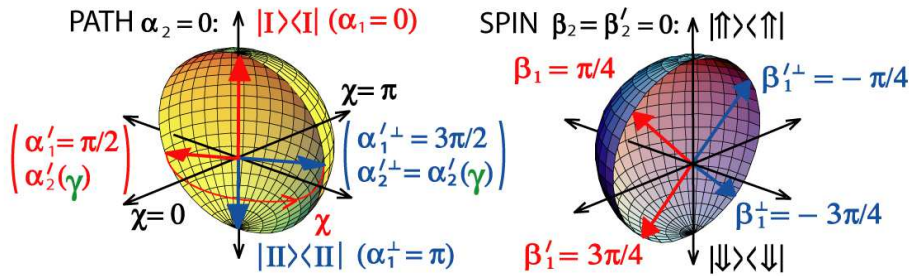}} \caption{\label{fig:azz}
Bloch-sphere description includes the measurement settings of
$\boldsymbol\alpha$ and $\boldsymbol\beta(\delta)$, determining the
projection operators, used for joint measurement of spin and path.}
\end{center}
\end{figure}

Next we discuss the situation where the standard maximal value $S=2\sqrt{2}$ can be achieved by keeping the polar angles $\alpha'_1$, $\beta_1$
and $\beta'_1$ constant at the Bell angles $\alpha'_1=\frac{\pi}{2}$, $\beta_1=\frac{\pi}{4}$, $\beta'_1=\frac{3\pi}{4}$, ($\alpha_1=0$), while
the azimuthal parts, $\alpha'_2$, $\beta_2$ and $\beta'_2$ ($\alpha_2=0$), are varied. A Bloch sphere description of this configuration can be seen in
Fig.\,\ref{fig:azz}. The corresponding S function is denoted as
\begin{eqnarray}
    \fl S(\alpha'_2,\beta_2,\beta'_2,\gamma)
    = \Bigl\lvert-\sqrt{2}-\frac{\sqrt{2}}{2}\Bigl(\cos(\alpha'_2-\beta_2-\gamma)+\cos(\alpha'_2-\beta'_2-\gamma)\Bigr)\Bigr\rvert\;.
\end{eqnarray}
The maximum value $2\sqrt{2}$ is reached for
\begin{eqnarray}\label{eq:azzMax}
\beta_2=\beta'_2\textrm{, }\label{eq:Cond3}\quad\rm{and}\quad
\alpha'_2-\beta'_2&=&\gamma\, (\textrm{mod}\,\pi)\label{eq:Cond4}.
\end{eqnarray}
For convenience $\beta_2=0$ is chosen.

\begin{figure}[t]
\begin{center}
\scalebox{0.31}{\includegraphics{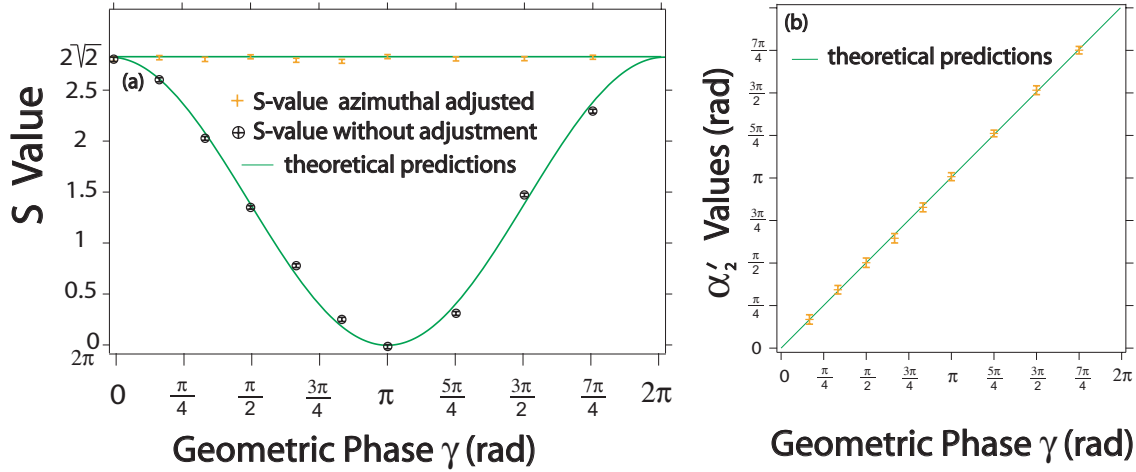}}
\caption{\label{fig:ResultsAzimuthAdjust} (Color online) (a)
\emph{Azimuthal adjusted} $S$-values versus geometric phase $\gamma$
with balanced Bell angle ($\alpha'_2$) according to the geometric
phase $\gamma$, and without corrections. (b) the corresponding
modified Bell angle is plotted versus the geometric phase $\gamma$.}
\end{center}
\end{figure}

Experimentally the angle between the measurement planes is adjusted
by one azimuthal angle ($\alpha'_2$), which is deduced by phase
shifter ($\chi$) scans.

For the spin measurement the directions are fixed and given by
$\boldsymbol\beta$: $\beta_1=\pi/4$, $\beta_2=0$ and
$\boldsymbol\beta'$: $\beta'_1=3\pi/4$, $\beta'_2=0$ (together with
$\beta^\bot_1=-3\pi/4$, $\beta'^{\bot}_1 =-\pi/4$. For the projective path
measurement the fixed directions read as $\alpha_1=0$
($\alpha_1^\bot=\pi$),  for measurements with beam block, and
$\alpha'_1=\pi/2$ ($\alpha'^\bot_1=3\pi/2$). Phase shifter ($\chi$)
scans are performed in order to determine $\alpha'_2$ for a maximal
violation of the Bell-like CHSH inequality yielding $S=2\sqrt2$.

As predicted by Eq.(\ref{eq:azzMax}) the constant maximal $S$ value of $2\sqrt{2}$ (see Fig.\,\ref{fig:ResultsAzimuthAdjust}\,(a))
is found for $\alpha'_2=\gamma$, which is plotted in
Fig.\,\ref{fig:ResultsAzimuthAdjust}\,(b). In Fig.\,\ref{fig:ResultsAzimuthAdjust}\,(a), the case is
included where no corrections are applied to the Bell angles.
According to Eq.(\ref{eq:Sunchganed}) the familiar maximum value of
$2\sqrt{2}$ is reached only for $\gamma=0$, and at $\gamma=\pi$ the
value of $S=0$ is found.

This experiment demonstrates in particular, that a geometric phase in one subspace does not lead to a loss of entanglement. Two schemes, polar and azimuthal adjustment of the Bell
angles, are realized, balancing the influence of the geometric phase. The former scheme yields a sinusoidal oscillation of the correlation function $S$, such that it varies in the range between 2 and $2\sqrt{2}$ and, therefore, always exceeds the boundary value 2 between quantum mechanical and noncontextual hidden-variable theories. The latter scheme results in a constant, maximal violation of the Bell-like CHSH inequality, where $S$ remains $2\sqrt2$ independent of the value of the geometric phase $\gamma$.

\section{Concluding remarks}

Neutron optical experiments are used for studying characteristics of phases of geometric origin.  First, non-additivity of the mixed state phase has been observed in a polarimetric experiment. Since the purity of quantum states in real experiments is always smaller than 1, non-additivity is of importance in all applications of quantum phases. Thinking about phase gates, it means that the purity of the utilized quantum system has to be considered when inducing phases for further processing. Second, a technique for geometric phase generation has been established by means of a precise spin manipulation due to interaction with rf-fields, in an interferometric setup. Applying the formalism of the Jaynes-Cumming Hamiltonian to the patterns in the observed outgoing beam of the interferometer, we find good agreement between experiment and theory. This technique is also applied to phase manipulations of the spin subspace in a triple-entanglement experiment with neutrons, which will be topic of a forthcoming publication. Finally, the effect of the geometric phase on the entanglement of the system, has been analyzed in detail, using a Bell-like CHSH inequality. It is demonstrated, how the effects of the geometric phase on the outcome of a Bell measurement can be balanced by an appropriate change of Bell angles. Neutrons have proved to be a suitable quantum system for studying topological effects. Interferometric as well as polarimetric techniques will lead to further investigations, relevant for possible applications
of the geometric phase. For instance, we are planing a polarimetric experiment, in which the geometric phase for non-unitary evolutions is considered.

\subsection{Acknowledgments}
This work has been supported by the Austrian Science Foundation, FWF (P21193-N20, P-17803-N02 and P-
20265). K.D.-R. would like to thank the FWF for funding her work by a Hertha Firnberg Position (T389-N16).


\newpage

\begin{thebibliography}{10}

\bibitem{BerryProcRSocLondA1984}
M.~V. Berry.
\newblock Quantal phase factors accompanying adiabatic changes.
\newblock {\em Proc. R. Soc. Lond. A}, 392:45, 1984.

\bibitem{TomitaPRL1986}
A.~Tomita and R.~Y. Chiao.
\newblock Observation of Berry's topological phase by use of an optical fiber.
\newblock {\em Phys. Rev. Lett.}, 57:937, 1986.

\bibitem{BitterPRL1987}
T.~Bitter and D.~Dubbers.
\newblock Manifestation of Berry\char39{}s topological phase in neutron spin rotation.
\newblock {\em Phys. Rev. Lett.}, 59:251, 1987.

\bibitem{AharonovPRL1987}
Y.~Aharonov and J.~S. Anandan.
\newblock Phase change during a cyclic quantum evolution.
\newblock {\em Phys. Rev. Lett.}, 58:1593, 1987.

\bibitem{SamuelPRL1988}
J.~Samuel and R.~Bhandari.
\newblock General setting for Berry's phase.
\newblock {\em Phys. Rev. Lett.}, 60:2339, 1988.

\bibitem{ManiniPRL2000}
N.~Manini and F.~Pistolesi.
\newblock Off-diagonal geometric phases.
\newblock {\em Phys. Rev. Lett.}, 85:3067, 2000.

\bibitem{UhlmannLettMathPhys1991}
A.~Uhlmann.
\newblock A gauge field governing parallel transport along mixed states.
\newblock {\em Lett. Math. Phys.}, 21:229, 1991.

\bibitem{SjoeqvistPRL2000}
E.~Sj{\"o}qvist, A.~K. Pati, A.~Ekert, J.~S. Anandan, M.~Ericsson, D.~K.~L. Oi, and
  V.~Vedral.
\newblock Geometric phases for mixed states in interferometry.
\newblock {\em Phys. Rev. Lett.}, 85:2845, 2000.

\bibitem{DuPRL2003}
J.~Du, P.~Zou, M.~Shi, L.~C. Kwek, J.~W. Pan, C.~H. Oh, A.~Ekert, D.~K.~L. Oi, and
  M.~Ericsson.
\newblock Observation of geometric phases for mixed states using {NMR}
  interferometry.
\newblock {\em Phys. Rev. Lett.}, 91:100403, 2003.

\bibitem{EricssonPRL2005}
M.~Ericsson, D.~Achilles, J.~T. Barreiro, D.~Branning, N.~A. Peters, and P.~G. Kwiat.
\newblock Measurement of geometric phase for mixed states using single photon
  interferometry.
\newblock {\em Phys. Rev. Lett.}, 94:050401, 2005.

\bibitem{FilippPRL2003}
S.~Filipp and E.~Sj{\"o}qvist.
\newblock Off-diagonal geometric phase for mixed states.
\newblock {\em Phys. Rev. Lett.}, 90:050403, 2003.

\bibitem{FilippPRA2003}
S.~Filipp and E.~Sj{\"o}qvist.
\newblock Off-diagonal generalization of the mixed-state geometric phase.
\newblock {\em Phys. Rev. A}, 68:042112, 2003.

\bibitem{RauchWerner2000}
H.~Rauch and S.~A. Werner.
\newblock {\em Neutron Interferometry}.
\newblock Clarendon Press, Oxford, UK, 2000.

\bibitem{Rauch-pla74}
H.~Rauch, W.~Treimer, and U.~Bonse.
\newblock Test of a single crystal neutron interferometer.
\newblock 47:369, 1974.

\bibitem{hasegawaPRA1996}
Y.~Hasegawa, M.~Zawisky, H.~Rauch, and A.~I. Ioffe.
\newblock Geometric phase in coupled neutron interference loops.
\newblock {\em Phys. Rev. A}, 53:2486, 1996.

\bibitem{FilippPRA2005}
S.~Filipp, Y.~Hasegawa, R.~Loidl, and H.~Rauch.
\newblock Noncyclic geometric phase due to spatial evolution in a neutron
  interferometer.
\newblock {\em Phys. Rev. A}, 72:021602(R), 2005.

\bibitem{waghPRL1998}
A.~G. Wagh, V.~C. Rakhecha, P.~Fischer, and  A.~I. Ioffe.
\newblock Neutron interferometric observation of noncyclic phase.
\newblock {\em Phys. Rev. Lett.}, 81:1992, 1998.

\bibitem{Allman-pra97}
B.~E. Allman, H.~Kaiser, S.~A. Werner, A.~G. Wagh, V.~C. Rakhecha, and J.~Summhammer.
\newblock Observation of geometric and dynamical phases by neutron
  interferometry.
\newblock 56:4420, 1997.

\bibitem{NielsenCuang2000}
M.~A. Nielsen and I.~L. Chuang.
\newblock {\em Quantum Computation and Quantum Information}.
\newblock Cambridge Unviversity Press, Cambridge, 2000.

\bibitem{LeekScience2007}
P.~J. Leek, J.~M. Fink, A.~Blais, R.~Bianchetti, M.~G{\"o}ppl, J.~M. Gambetta, D.~I.
  Schuster, L.~Frunzio, R.~J. Schoelkopf, and A.~Wallraff.
\newblock Observation of Berry's phase in a solid-state qubit.
\newblock {\em Science}, 318:1889, 2007.

\bibitem{FilippPRL2009}
S.~Filipp, J.~Klepp, Y.~Hasegawa, C.~Plonka-Spehr, U.~Schmidt, P.~Geltenbort, and
  H.~Rauch.
\newblock Experimental demonstration of the stability of Berry's phase for a
  spin-1/2 particle.
\newblock {\em Phys. Rev. Lett.}, 102:030404, 2009.

\bibitem{BertlmannPRA2004}
R.~A. Bertlmann, K.~Durstberger, Y.~Hasegawa, and B.~C. Hiesmayr.
\newblock Berry phase in entangled systems: A proposed experiment with single
  neutrons.
\newblock {\em Phys. Rev. A}, 69:032112, 2004.

\bibitem{SjoeqvistPRA2000}
E.~Sj{\"o}qvist.
\newblock Geometric phase for entangled spin pairs.
\newblock {\em Phys. Rev. A}, 62:022109, 2000.

\bibitem{tongJPhysA2003}
D.~M. Tong, L.~C. Kwek, and C.~H. Oh.
\newblock Geometric phase for entangled states of two spin 1/2 particles in
  rotating magnetic field.
\newblock {\em J. Phys. A}, 36(2):1149, 2003.

\bibitem{HasegawaNature2003}
Y.~Hasegawa, R.~Loidl, G.~Badurek, M.~Baron, and H.~Rauch.
\newblock Violation of a Bell-like inequality in single-neutron interferometry.
\newblock {\em Nature (London)}, 425:45, 2003.

\bibitem{HasegawaPRA2007}
Y.~Hasegawa, R.~Loidl, G.~Badurek, S.~Filipp, J.~Klepp, and H.~Rauch.
\newblock Evidence for entanglement and full tomographic analysis of bell
  states in a single-neutron system.
\newblock {\em Phys. Rev. A}, 76:052108, 2007.

\bibitem{KleppPLA2005}
J.~Klepp, S.~Sponar, Y.~Hasegawa, E.~Jericha, and G.~Badurek.
\newblock Noncyclic pancharatnam phase for mixed state {SU(2)} evolution in
  neutron polarimetry.
\newblock {\em Phys. Lett. A}, 342:48, 2005.

\bibitem{SinghPRA2003}
K.~Singh, D.~M. Tong, K.~Basu, J.~L. Chen, and J.~F. Du.
\newblock Geometric phases for nondegenerate and degenerate mixed states.
\newblock {\em Phys. Rev. A}, 67:032106, 2003.

\bibitem{KleppPRL2008}
J.~Klepp, S.~Sponar, S.~Filipp, M.~Lettner, G.~Badurek, and Y.~Hasegawa.
\newblock Observation of nonadditive mixed-state phases with polarized
  neutrons.
\newblock {\em Phys. Rev. Lett.}, 101:150404, 2008.

\bibitem{KleppAIPProc2009}
J.~Klepp, S.~Sponar, S.~Filipp, M.~Lettner, G.~Badurek, and Y.~Hasegawa.
\newblock Nonadditive mixed state phases in neutron optics.
\newblock In {\em Foundations of Probability and Physics - {5}}, page 314,
  Melville, New York, USA, 2009. American Institute of Physics.

\bibitem{Pancharatnam1956}
S.~Pancharatnam.
\newblock Generalized theory of interference, and its applications.
\newblock {\em Proc. Indian Acad. Sci.}, 44:247, 1956.

\bibitem{WaghPLA1995}
A.~G. Wagh and V.~C. Rakhecha.
\newblock On measuring the Pancharatnam phase.
II. SU (2) polarimetry.
\newblock {\em Phys. Lett. A}, 197:112, 1995.

\bibitem{LarssonPLA2003}
P.~Larsson and E.~Sj{\"o}qvist.
\newblock Noncyclic mixed state phase in {SU(2) polarimetry}.
\newblock {\em Phys. Lett. A}, 315:12, 2003.

\bibitem{BertlmannPRA2006}
R.~A. Bertlmann, K.~Durstberger, and Y.~Hasegawa.
\newblock Decoherence modes of entangled qubits within neutron interferometry.
\newblock {\em Phys. Rev. A}, 73:022111, 2006.

\bibitem{JaynesProcIEEE1963}
E.~T. Jaynes and F.~W. Cummings.
\newblock Comparison of quantum and semiclassical radiation theories with
  application to the beam maser.
\newblock {\em Proc. IEEE}, 51:89, 1963.

\bibitem{ShoreJModOpt1993}
B.~W. Shore and P.~L. Knight.
\newblock Topical review of the Jaynes-Cummings model.
\newblock {\em J. Mod. Optics}, 40:1195, 1993.

\bibitem{MuskatPRL1987}
E.~Muskat, D.~Dubbers, and O.~Sch{\"a}rpf.
\newblock Dressed neutrons.
\newblock {\em Phys. Rev. Lett.}, 58:2047, 1987.

\bibitem{CohenTannJPhys1969}
C.~Cohen-Tannoudji and S.~Haroche.
\newblock Interpr\'etation quantique des diverses r\'esonances observ\'ees lors
  de la diffusion de photons optiques et de radiofr\'equence par un atome.
\newblock {\em J. Phys. (Paris)}, 30:125, 1969.

\bibitem{alefeldZPhysB1981}
B.~Alefeld, G.~Badurek, and H.~Rauch.
\newblock Observation of the neutron magnetic resonance energy shift.
\newblock {\em Z. Phys. B}, 41:231, 1981.

\bibitem{GaehlerPLA1987}
R.~G{\"a}hler and R.~Golub.
\newblock A neutron resonance spin echo spectrometer for quasi-elastic and
  inelastic scattering.
\newblock {\em Phys. Lett. A}, 43:123, 1987.

\bibitem{SummhammerPRA1993}
J.~Summhammer.
\newblock Coherent multiphoton exchange between a neutron and an oscillating
  magnetic field.
\newblock {\em Phys. Rev. A}, 47:556, 1993.

\bibitem{GolubAmJPhys1994}
R.~Golub, R.~G{\"a}hler, and T~Keller.
\newblock A plane wave approach to particle beam magnetic resonance.
\newblock {\em Am. J. Phys.}, 62:9, 1994.

\bibitem{SummhammerPRL1995}
J.~Summhammer, K.~A. Hamacher, H.~Kaiser, H.~Weinfurter, D.~L. Jacobson, and S.~A.
  Werner.
\newblock Multiphoton exchange amplitudes observed by neutron interferometry.
\newblock {\em Phys. Rev. Lett.}, 75:3206, 1995.

\bibitem{GrigorievPRA2004}
S.~V. Grigoriev, W.~H. Kraan, and M.~T. Rekveldt.
\newblock Four-wave neutron-resonance spin echo.
\newblock {\em Phys. Rev. A}, 69:043615, 2004.

\bibitem{SponarPRA2008}
S.~Sponar, J.~Klepp, R.~Loidl, S.~Filipp, G.~Badurek, Y.~Hasegawa, and H.~Rauch.
\newblock Coherent energy manipulation in single-neutron interferometry.
\newblock {\em Phys. Rev. A}, 78:061604(R), 2008.

\bibitem{SuterPRL1988}
D.~Suter, K.~T. Mueller, and A.~Pines.
\newblock Study of the Aharonov-Anandan quantum phase by NMR interferometry.
\newblock {\em Phys. Rev. Lett.}, 60:1218, 1988.

\bibitem{BlochPR1940}
F.~Bloch and A.~Siegert.
\newblock Magnetic resonance for nonrotating fields.
\newblock {\em Phys. Rev.}, 57(6):522, 1940.

\bibitem{BellPhys1964}
J.~S. Bell.
\newblock On the Einstein-Podolsky-Rosen paradox.
\newblock {\em Physics (Long Island City, N.Y.)}, 1(3):195, 1964.

\bibitem{ClauserPRL1969}
J.~F. Clauser, M.~A. Horne, A.~Shimony, and R.~A. Holt.
\newblock Proposed experiment to test local hidden-variable theories.
\newblock {\em Phys. Rev. Lett.}, 23(15):880, 1969.

\bibitem{BasuPLA2001}
S.~Basu, S.~Bandyopadhyay, G.~Kar, and D.~Home.
\newblock Bell's inequality for a single spin-1/2 particle and quantum
  contextuality.
\newblock {\em Phys. Lett. A}, 279:281, 2001.

\end{thebibliography}
\end{document}